\documentclass[conference]{IEEEtran}
\usepackage{amsfonts}
\usepackage{amssymb}
\usepackage{cite}
\usepackage[cmex10]{amsmath}
\usepackage{float}
\usepackage{color}
\usepackage{stfloats,fancyhdr}
\usepackage{amsmath}
\usepackage{algorithm}
\usepackage{algorithmic}
\usepackage{multirow}
\usepackage{changepage}
\usepackage[normalem]{ulem}

\newtheorem{remark}{Remark}

\IEEEoverridecommandlockouts

\ifCLASSINFOpdf
  \usepackage[pdftex]{graphicx}
  \DeclareGraphicsExtensions{.pdf,.jpeg,.png}
\else
  \usepackage[dvips]{graphicx}
  \DeclareGraphicsExtensions{.eps}
\fi

\usepackage{subfigure}

\usepackage{fancybox,dashbox}

\begin{document}
%
\title{{\huge A Discrete Time-Switching Protocol for Wireless-Powered Communications with Energy Accumulation}}
\author{Yifan Gu, He~(Henry)~Chen, Yonghui Li, and Branka Vucetic
\\
School of Electrical and Information Engineering, The University of Sydney, Sydney, NSW 2006, Australia\\
Email: yigu6254@uni.sydney.edu.au, \{he.chen, yonghui.li, branka.vucetic\}@sydney.edu.au
}

\maketitle

\begin{abstract}
This paper investigates a wireless-powered communication network (WPCN) setup with one multi-antenna access point (AP) and one single-antenna source. It is assumed that the AP is connected to an external power supply, while the source does not have an embedded energy supply. But the source could harvest energy from radio frequency (RF) signals sent by the AP and store it for future information transmission. We develop a discrete time-switching (DTS) protocol for the considered WPCN. In the proposed protocol, either energy harvesting (EH) or information transmission (IT) operation is performed during each transmission block. Specifically, based on the channel state information (CSI) between source and AP, the source can determine the minimum energy required for an outage-free IT operation. If the residual energy of the source is sufficient, the source will start the IT phase. Otherwise, EH phase is invoked and the source accumulates the harvested energy. To characterize the performance of the proposed protocol, we adopt a discrete Markov chain (MC) to model the energy accumulation process at the source battery. A closed-form expression for the average throughput of the DTS protocol is derived. Numerical results validate our theoretical analysis and show that the proposed DTS protocol considerably outperforms the existing harvest-then-transmit protocol when the battery capacity at the source is large.
\end{abstract}


\IEEEpeerreviewmaketitle
\section{Introduction}
The performance of many wireless networks is largely confined by the energy constrained nodes that requires replenishment periodically. Conventionally, energy harvesting from ambient environment, such as solar power and wind energy, was treated as a sustainable solution to prolong the lifetime of such networks. Recently, a novel radio-frequency (RF) energy harvesting technique has emerged and drawn a lot of interest (see \cite{RFEH} and the references there in), in which the RF signals are treated as a new viable source for energy harvesting. In \cite{DC1} and \cite{DC2}, it is shown that through appropriate circuits, devices are able to capture the energy carried by the RF signals and convert them into a direct current. Compared to the conventional energy harvesting techniques, RF energy harvesting can power the nodes from a radio environment and has the feature of a low-power and long-distance transfer. As a result, RF energy harvesting technique has found its applications widely in wireless sensor systems \cite{sensor} and wireless body systems \cite{body}. Along this new energy harvesting technique, a new type of wireless networks named wireless-powered communication network (WPCN) has emerged and attracted significant research interests. In the WPCN, wireless nodes are not equipped with embedded electric supplies and their operations rely only on harvesting the energy from a radio environment.

The authors in \cite{harvest-then-transmit} investigated a particular WPCN consisting of one hybrid access point (AP) and multiple users. A harvest-then-transmit (HTT) protocol was proposed. In the HTT protocol, during each transmission block, the users first harvest energy from the RF signals broadcasted by the AP in the downlink (DL). After that, the users perform uplink (UL) information transmission in orthogonal time slots using the harvested energy. \cite{extend} extended the works in \cite{harvest-then-transmit} and studied the network setup with one multi-antenna AP and many single-antenna users. After energy harvesting phase, the users can transmit information simultaneously to the AP through space-division-multiple-access technique. Moreover, \cite{Chen_TSP_2015_Harvest} and \cite{Gu_ICC_2015} focused on the design and performance analysis of cooperative protocols for WPCNs with different setups. Very recently in \cite{optimal}, the authors studied a WPCN equipped with one multi-antenna AP and one single-antenna user. Considering an HTT protocol implemented in the network, the optimal energy harvesting time was derived for both delay-limited transmission and delay-tolerant transmission modes. Among these works \cite{harvest-then-transmit, extend, optimal,Chen_TSP_2015_Harvest,Gu_ICC_2015}, energy harvesting is always followed by information transmission operation within each transmission block and the throughput of the system is maximized by finding the optimal energy harvesting time.

However, the energy accumulation and long-term energy harvesting (EH)/information transmission (IT) scheduling across different transmission blocks has not been considered so far. In particular, the harvested energy in the current transmission block could be saved to transmit information in the following transmission blocks, or the block with good UL channel condition can be used to transmit information entirely. Therefore, with the properly designed EH and IT scheduling and energy accumulation, it is expected that the system throughput can be further improved. Motivated by this, in this paper we develop a discrete time-switching framework for the WPCNs. Specifically, we investigate a WPCN setup with one multi-antenna AP and one single-antenna source\footnote{We assume that source is a low-energy and low-cost device such that it could not be equipped with multiple antenna due to cost/size constraints.}. A discrete time-switching protocol is proposed for the considered network, in which the source only performs either energy harvesting (EH) or information transmission (IT) operation in each entire transmission block. We assume that full channel state information (CSI) is available at both AP and source sides for optimal energy beamforming and transmission scheduling. At the beginning of each block, the source first checks whether the current residual energy is enough to perform an outage-free UL information transmission. If yes, the  source will transmit information to the AP in the current block with the minimum required energy that guarantees an outage-free transmission. Otherwise, the  AP will perform DL energy transfer to the source and the source stores the harvested energy for future usage. To our best knowledge, this is the first paper that designs and analyzes the discrete time-switching strategy for WPCNs with energy accumulation and adaptive transmit power at the source. Note that in contrast to the existing HTT protocol, the source does not need to exhaust it harvested energy during each IT block in the proposed DTS protocol. Thus, we need to characterize the complicated energy accumulation process of the source battery, which makes the performance analysis of the proposed DTS protocol challenging.

The main contributions of this paper can be summarized as follows:
\begin{itemize}
\item  We propose a discrete time-switching (DTS) protocol for the considered WPCN, in which the EH and IT operations are switched across transmission blocks. This is contrast to the existing HTT protocol that splits each transmission block to perform both EH and IT, leading to higher implementation complexity than the proposed one.
\item To characterize the average throughput of the proposed DTS protocol, we adopt a discrete Markov Chain (MC) \cite{MC} to model the energy accumulation process of the source battery. Based on this, a closed-form expression for the average throughput of the DTS protocol is then derived over Rayleigh fading channels.
\item Numerical results validate all the theoretical analysis, examine the impact of various system parameters, and compare the throughput performance of DTS protocol with the existing HTT protocol with its optimal time allocation. It is shown that the proposed DTS protocol outperforms the HTT one when the battery capacity at the source is large.
\end{itemize}
\begin{figure}
\centering \scalebox{0.55}{\includegraphics{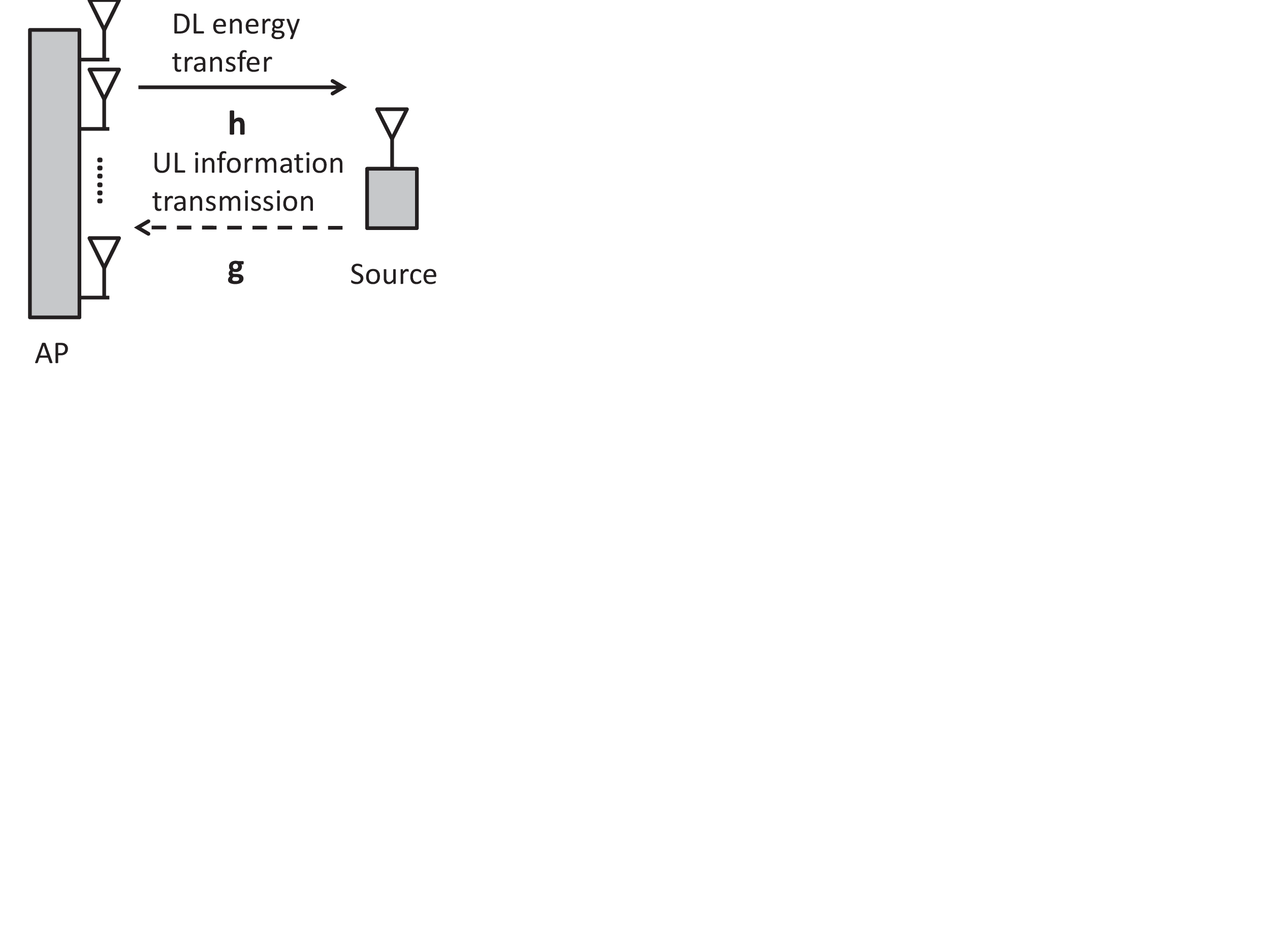}}
\caption{System model of the considered multi-antenna WPCN. \label{fig:WPCN}}
\end{figure}
\section{System Model and DTS Protocol}
As shown in Fig.\ref{fig:WPCN}, we consider a multiple-antenna WPCN consists of one multi-antenna AP equipped with $N>1$ antennas and one single-antenna source\footnote{The extension to the multi-user scenario has been considered as our future work.}. We assume that the hybrid AP is connected to an external energy supply, while the source has no embedded electric supply but is only equipped with a discrete-level battery with finite size. Thus, the source can harvest energy from the RF signals broadcasted by the AP in the DL and store the harvested energy for further information transmission in the UL.

The communication is performed in blocks with the duration $T$. We assume that the channels between the source and AP experience slow, independent, and frequency-flat fading such that the channel power gain remains unchanged within each block but changes independently from one block to the other. The block diagram of the proposed DTS protocol is shown in Fig. 2(a). As a comparison, the block diagram of the HTT \cite{harvest-then-transmit} protocol is also illustrated in Fig. 2(b). In the HTT protocol, EH and IT operations are time switched within each transmission block. And there exists an optimal time allocation between EH and IT in each block \cite{optimal}. In contrast, EH and IT operations are switched for each discrete transmission block in the proposed DTS protocol, which may achieve a lower implementation complexity relative to the HTT protocol. We also assume that the full channel state information (CSI) is available at both the AP and the source sides for optimal energy beamforming and transmission scheduling\footnote{This channel information can be obtained via channel estimation technique at the beginning of each transmission block. For the purpose of exploration, we assume that the time duration for this channel estimation process is negligible compared with the whole block duration.}. Thus, the source can determine the minimum energy required to guarantee that no outage happens in the IT operation. If the source has sufficient energy, the source will start the IT phase. Otherwise, EH operation is invoked and the source stores the energy harvested from the AP. In the following, let's first characterize the minimum energy required for a successive IT and the amount of energy harvested in a EH block for a given channel realization.
\begin{figure}
\centering
 \subfigure[DTS protocol]
  {\scalebox{0.45}{\includegraphics {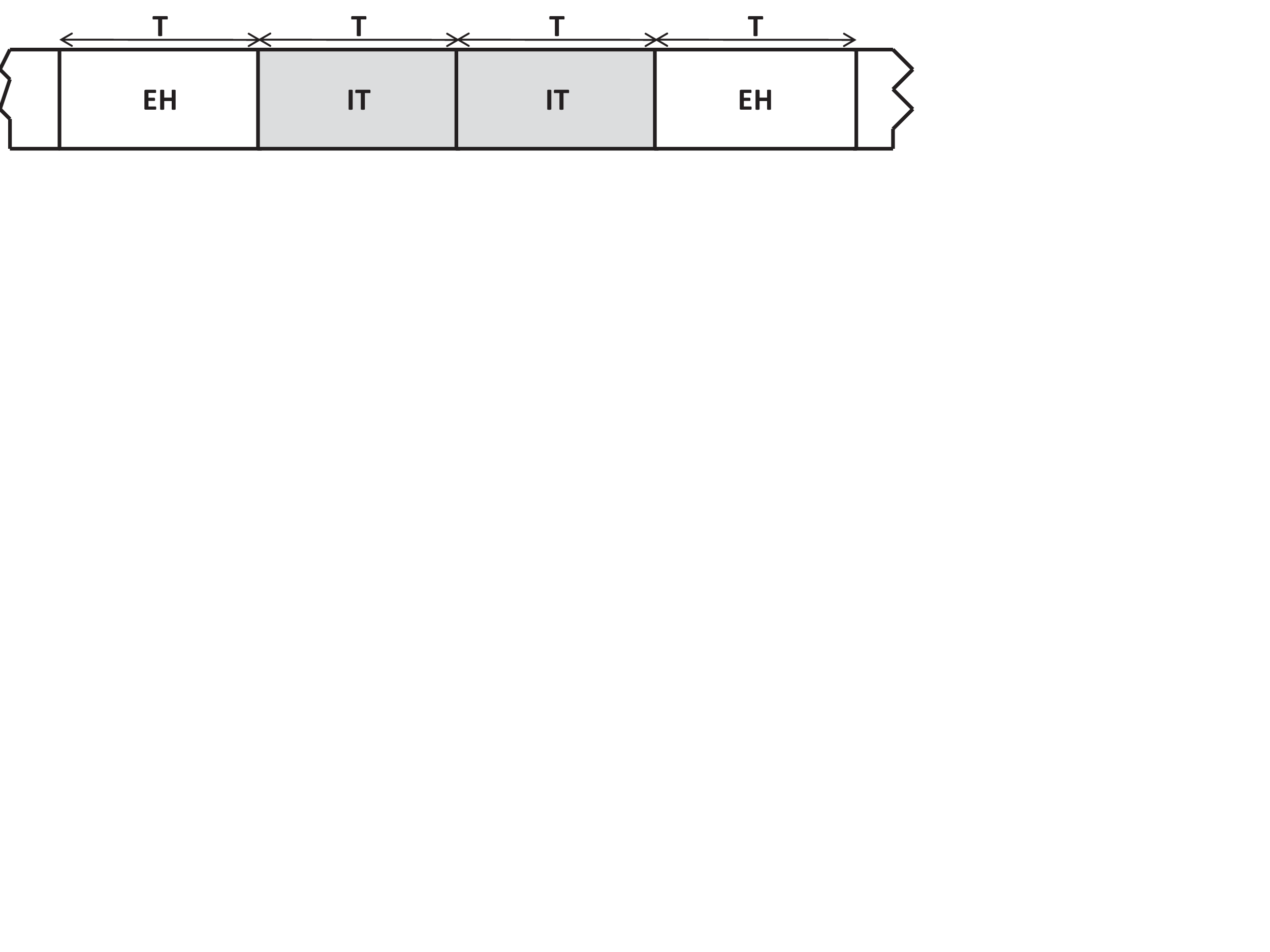}
  \label{fig:dts}}}
\hfil
 \subfigure[HTT protocol]
  {\scalebox{0.45}{\includegraphics {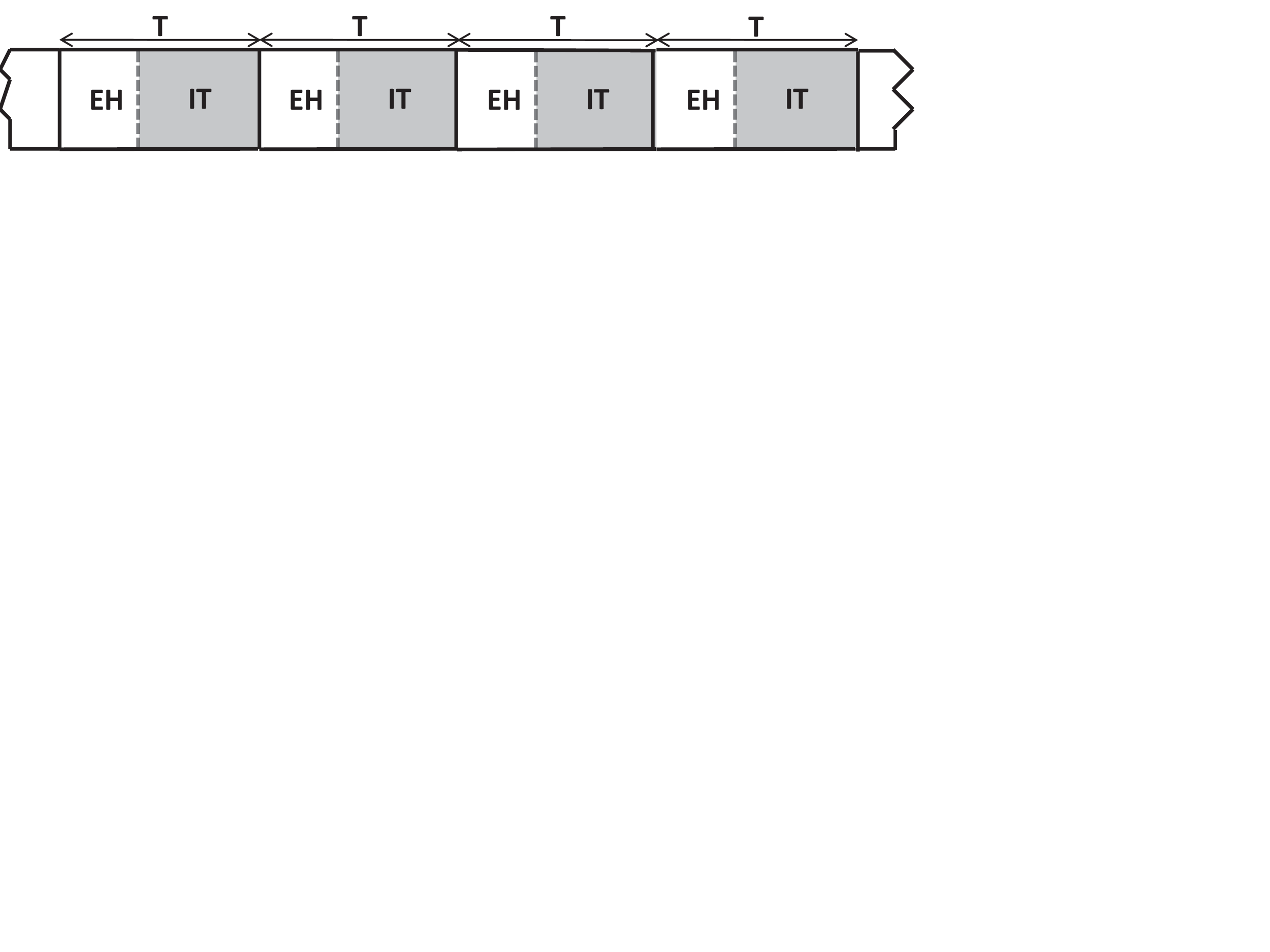}
\label{fig:HTT}}}
\caption{Block diagram of the discrete time-switching protocol and harvest-then-transmit protocol.
\label{fig:time diagram}}
\end{figure}

To this end, we use $\textbf{h}$ and $\textbf{g}$ to denote the DL and UL channel gain vectors of the current transmission block with a dimension $N$ and assume that their entries follow independent and identical distributed (i.i.d) circularly symmetric complex Gaussian distributions with a zero mean and variance of $\Omega$. Let $P$ denote the transmit power of the hybrid AP. For simplicity, we consider a normalized transmission block (i.e., $T=1$) hereafter. When the source chooses to perform EH in the current block, the maximal ratio transmission (MRT) with beamforming vector $ {\bf{w}}={\bf{h}}/\left\|{\bf{h}}\right\|$ is known to be optimal for DL energy transfer \cite{MRT}. The harvested energy at the source can thus be expressed as
\begin{equation}\label{EH}
{E_H}{ = }\eta  P\left\| \mathbf{h} \right\|^2,
\end{equation}
by ignoring the negligible energy harvested from the noise. In (\ref{EH}), $\eta$ is the energy conversion efficiency and $\left\|   \cdot  \right\|$ denotes the norm of a given vector.

On the other hand, the source can also decide to perform IT during the current transmission block. In this case, the maximum ratio combination (MRC) technique is implemented at the AP to maximize the received signal-to-noise ratio (SNR). Let $R$ denote the fixed bit rate of the source. Then, the minimum required energy $E_T$ to guarantee an outage-free transmission in the UL should satisfy
\begin{equation}
{\log _2}\left( {1 + {{{E_T}{{\left\| \mathbf{g} \right\|}^2}} \over {{N_0}}}} \right) = R,
\end{equation}
where ${N_0}$ is the variance of additive white Gaussian noise (AWGN) at the AP.
After some simple manipulation, we have
\begin{equation}\label{transmitenergy}
{E_T} = {{\upsilon {N_0}} \over {{{\left\| \mathbf{g} \right\|}^2}}},
\end{equation}
where $\upsilon  = {2^R} - 1$.

In this paper, we consider a practical discrete-level model for the source battery \cite{row-stac}. Let $C$ denote the capacity of the source battery and $L$ denote the number of discrete levels. Then, the $i$th energy level of the source battery can be characterized as ${\varepsilon_i} = iC/L$, $i  \in \left\{  0,1,2 \cdots L\right\}$. We define state ${S_i}$ as the source residual battery being ${\varepsilon_i}$. The transition probability $T_{i,j}$ is defined as the probability of transition from state $S_i$ to state $S_j$. With this discrete battery model, the amount of harvested energy for a EH block  given in (1) should be discretized. This means the effective (discrete) harvested energy at the source is given by
\begin{equation}
{\varepsilon _H} \buildrel \Delta \over = {\varepsilon _{j}} \quad \text{where} \quad {j} = \arg \mathop {\max }\limits_{i \in \left\{ {0,1, \cdots ,L} \right\}} \bigg\{{\varepsilon _i}:{\varepsilon _i} < {E_H}\bigg\},
\end{equation}
Similarly, the actual amount of energy (discrete energy) used for a IT block can be expressed as

\begin{equation}\label{yifan}
{\varepsilon _T} \buildrel \Delta \over = {\varepsilon _{j}} \quad \text{where} \quad j= \arg \mathop {\min }\limits_{i \in \left\{ {0,1, \cdots ,L} \right\}} \bigg\{{\varepsilon _i}:{\varepsilon _i} > {E_T}\bigg\}.
\end{equation}

We are now ready to describe the behaviors of the proposed DTS protocol mathematically. Let $\zeta \left[ m \right] \in \left\{ {{\zeta _H},{\zeta _T}} \right\}$ denote the operation mode of the source for the $m$-th transmission block, where ${\zeta _H}$ and ${\zeta _T}$ denote the EH and IT modes, respectively. Let ${\varepsilon _R}[m]$ denote the residual energy stored at the source and $\Phi \left[ m \right]$ denote the throughput of the system for the $m$-th transmission block. Then, based on the principle of the proposed DTS protocol, we can readily have the following equations
\begin{equation}\label{indicator}
\zeta \left[ m \right] = \left\{ {
\begin{matrix}
   {{\zeta _H},\quad \text{if}\quad {\varepsilon _R}\left[ m \right] < {\varepsilon _T}}  \\
   {{\zeta _T},\quad \text{otherwise}}  \\
\end{matrix}
  } \right.
\end{equation}

\begin{equation}
{\varepsilon _R}\left[ m+1 \right] = \left\{ {
\begin{matrix}
   {\min \bigg \{{\varepsilon _R}\left[ {m } \right] + {\varepsilon _H}, C \bigg\}, \quad \text{if} \quad {\varepsilon _R}\left[ m \right] < {\varepsilon _T}}  \\
   {{\varepsilon _R}\left[ {m } \right] - {\varepsilon _T},\quad \text{otherwise}}  \\
\end{matrix}
  } \right.
\end{equation}

\begin{equation}\label{throughput}
\Phi \left[ m \right] = \left\{ {
\begin{matrix}
   {0,\quad \text{if}\quad {\varepsilon _R}\left[ m \right] < {\varepsilon _T}}  \\
   {R,\quad \text{otherwise}}  \\
\end{matrix}
  } \right.
\end{equation}
\section{Throughput analysis}
In this section, we derive a closed-form expression for the average throughput of the proposed DTS protocol. To this end, we first characterize the distribution of the source residual energy ${\varepsilon _R}$ by modeling the energy accumulation process at the source as a discrete Markov Chain (MC) \cite{MC}. For notation simplicity, we define $H={{\left\| \mathbf{h} \right\|}^2}$ and $G={{\left\| \mathbf{g} \right\|}^2}$.
\subsection{Markov Chain of the DTS Protocol}
Inspired by \cite{row-stac}, we can summarize the state transition of the MC for the source battery into the following eight general cases. Here, we use $T_{i,j}$ to denote the transition probability for the event that the MC transits from state $S_i$ to state $S_j$.
\subsubsection{The battery remains empty (${S_0}\quad to\quad {S_0}$)}
We consider that the transition starts with the state $S_0$, i.e., the source battery is empty. In this case, EH operation should be performed and the battery still remains empty after one EH block. This indicates that the energy harvested by the source ${\varepsilon _H}$ is equal to zero. From the definition given in (4), the condition ${E_H} < C/L$ should hold so that the harvested energy during the EH operation can be discretized to zero. Hence the transition probability of this case can be expressed as
\begin{equation}
\begin{split}
{T_{0,0}} & = \Pr \left\{ {{\varepsilon _H}}  =  0  \right\} = \Pr \left\{ {H \le {C  \over {\eta PL}}} \right\}\\
& =1 - \exp \left( { - {C \over \eta PL \Omega }} \right)\sum\limits_{n = 0}^{N - 1} {{{{{\left( {{C \over \eta PL\Omega }} \right)}^n}} \over {n!}}},
\end{split}
\end{equation}
where the last equation is obtained based on the fact that $H$ follows a chi-square distribution with $2N$ degree of freedom \cite{optimal}.
\subsubsection{The battery is fully charged (${S_0}\quad to\quad {S_L}$)}In this case, the harvested energy should be greater than the capacity of the source and the transition probability is then given by
\begin{equation}
\begin{split}
{T_{0,L}} &= \Pr \left\{ {{{\varepsilon _H}} > C } \right\} = \Pr \left\{ {H > {{C } \over {\eta P}}} \right\}\\
& = \exp \left( { - {C \over \eta P \Omega }} \right)\sum\limits_{n = 0}^{N - 1} {{{{{\left( {{C \over \eta P\Omega }} \right)}^n}} \over {n!}}}.
\end{split}
\end{equation}
\subsubsection{The battery is partially charged (${S_0}\quad to\quad {S_i}$ with $0 < i < L$)}
The discrete energy ${\varepsilon _H}$ should equal to $iC/L$, which indicates that $E_H$ falls between the level $i$ and $i+1$ of the source battery. The transition probability $T_{0,i}$ can thus be calculated as
\begin{equation}
\begin{split}
{T_{0,i}} &= \Pr \left\{ {{\varepsilon _H}=iC/L} \right\} \\
&= \Pr \left\{ {{{iC } \over {\eta PL}} < H \le {{(i + 1)C } \over {\eta PL}}} \right\} \\
&= \exp \left( { - {iC \over \eta P L \Omega }} \right)\sum\limits_{n = 0}^{N - 1} {{{{{\left( {{iC \over \eta P L\Omega }} \right)}^n}} \over {n!}}} - \\
& \quad \exp \left( { - {(i+1)C \over \eta P L \Omega }} \right)\sum\limits_{n = 0}^{N - 1} {{{{{\left( {{(i+1) C \over \eta P L \Omega }} \right)}^n}} \over {n!}}}.
\end{split}
\end{equation}
\subsubsection{The non-empty and non-full battery remains unchanged (${S_i}\quad to\quad {S_i}$ with $0 < i  < L$)}
From the definition of discretization given in (\ref{yifan}), the minimum transmitted energy in the IT operation can be characterized when $E_T$ falls between energy level zero and level one. In that case, we use one unit of energy for an outage-free IT operation. Hence we can conclude that IT operation happens only in the transitions where the source residual energy is reduced. As a result, we can deduce that the EH operation should be invoked in this considered case, which means that the require energy $\varepsilon_T$ is larger than the residual energy at the beginning of the block. Furthermore, the other condition is that the discrete harvested energy ${\varepsilon _H}$ should equal to $0$. Therefore, the corresponding transition probability can be characterized as
\begin{equation}
\begin{split}
{T_{i,i}} &=  \Pr \bigg \{ \big ({\varepsilon _T} > iC/L \big ) \cap \big( {\varepsilon _H} =0 \big) \bigg \} \\
& = \Pr \left\{ {G \le {{\upsilon {N_0}L} \over {iC }}} \right\}\Pr \left\{ {H \le {C  \over {\eta PL}}} \right\}\\
& = \left[ {1 - \exp \left( { - {\upsilon {N_0}L \over iC\Omega }} \right)\sum\limits_{n = 0}^{N - 1} {{{{{\left( {{\upsilon {N_0}L \over iC\Omega }} \right)}^n}} \over {n!}}}} \right] \times \\
& \quad  \left[{ 1 - \exp \left( { - {C \over \eta PL\Omega }} \right)\sum\limits_{n = 0}^{N - 1} {{{{{\left( {{C \over \eta PL\Omega }} \right)}^n}} \over {n!}}}}\right],
\end{split}
\end{equation}
where the last equations follows based on the assumption that $G$ has the same distribution with that of $H$.
\subsubsection{The non-empty battery is partially charged (${S_i}\quad to\quad {S_j}$ with $0 < i < j< L$)}EH phase is selected in the current transmission block and the discrete harvested energy is equal to $(j-i)C/L$, the transition probability can thus be calculated~as
\begin{equation}
\begin{split}
{T_{i,j}} &= \Pr \bigg \{ \big ( {\varepsilon _T} > iC/L \big) \cap \big( {\varepsilon _H}=(j - i)C/L \big) \bigg \} \\
&= \left[ {1 - \exp \left( { - {\upsilon {N_0}L \over iC\Omega }} \right)\sum\limits_{n = 0}^{N - 1} {{{{{\left( {{\upsilon {N_0}L \over iC\Omega }} \right)}^n}} \over {n!}}}} \right] \times \\
& \quad   \left[ {\exp \left( { {(j-i) C \over \eta P L \Omega }} \right)\sum\limits_{n = 0}^{N - 1} {{{{{\left( {{(j-i)C \over \eta P L\Omega }} \right)}^n}} \over {n!}}} - }\right.\\
& \quad \left. {\exp \left( { {(j-i+1)C \over \eta P L \Omega }} \right)\sum\limits_{n = 0}^{N - 1} {{{{{\left( {{(j-i+1) C \over \eta P L \Omega }} \right)}^n}} \over {n!}}} }\right].
\end{split}
\end{equation}

\subsubsection{The non-empty and non-full battery is fully charged (${S_i}\quad to\quad {S_L}$ with $0 < i < L$)}The transition probability can be evaluated as
\begin{equation}
\begin{split}
{T_{i,L}} &= \Pr \bigg \{ \big ( {\varepsilon _T} > iC/L  \big )\cap \big( {\varepsilon _H} > (L - i)C/L \big) \bigg \} \\
&= \left[ {1 - \exp \left( { - {\upsilon {N_0}L \over iC\Omega }} \right)\sum\limits_{n = 0}^{N - 1} {{{{{\left( {{\upsilon {N_0}L \over iC\Omega }} \right)}^n}} \over {n!}}}} \right] \times \\
& \quad \exp \left( {  {(L-i) C \over \eta P L \Omega }} \right)\sum\limits_{n = 0}^{N - 1} {{{{{\left( {{(L-i) C \over \eta P L \Omega }} \right)}^n}} \over {n!}}}.
\end{split}
\end{equation}

\subsubsection{The battery remains full (${S_L}\quad to\quad {S_L}$ )} In this case, EH phase is performed in the current transmission block and the harvested energy can be any value since the battery is already full at the beginning of the transition. The transition probability is thus given by

\begin{equation}
\begin{split}
{T_{L,L}} &= \Pr \left\{ {{\varepsilon _T} > C } \right\} \\
&= \left[ {1 - \exp \left( { - {\upsilon {N_0} \over C\Omega }} \right)\sum\limits_{n = 0}^{N - 1} {{{{{\left( {{\upsilon {N_0} \over C\Omega }} \right)}^n}} \over {n!}}}} \right].\\
\end{split}
\end{equation}
\subsubsection{The non-empty battery is discharged (${S_j}\quad to\quad {S_i}$ with $0 \le i <j \le L$)}
Since the stored energy is reduced during the transition, IT phase is chosen in the current transmission block and the discrete used energy ${\varepsilon _T}$ is equal to $(j-i)C/L$. Thus, the transition probability is given by
\begin{equation}
\begin{split}
{T_{j,i}} &= \Pr  \bigg \{ \big ({\varepsilon _T} \le jC/L \big) \cap \big(  {\varepsilon _T}= (j - i)C/L  \big) \bigg \}\\
&=\Pr \left \{{{\varepsilon _T}= (j - i)C/L }\right\}\\
&= \exp \left( { - {\upsilon {N_0} L \over (j-i)C\Omega }} \right)\sum\limits_{n = 0}^{N - 1} {{{{{\left( {{\upsilon {N_0} L\over (j-i)C\Omega }} \right)}^n}} \over {n!}}} -\\
& \quad \exp \left( { - {\upsilon {N_0}L \over (j-i-1)C\Omega }} \right)\sum\limits_{n = 0}^{N - 1} {{{{{\left( {{\upsilon {N_0} L \over (j-i-1)C\Omega }} \right)}^n}} \over {n!}}}.
\end{split}
\end{equation}

Let $\mathbf{Z}=({T_{i,j}})$ denote the state transition matrix of the MC for source battery with dimension $L+1$. With reference to \cite{row-stac}, we can easily verify that the MC transition matrix $\mathbf{Z}$ is irreducible and row stochastic. Thus there must exists a unique solution $\mathbf{\pi }$ that satisfies the following equation \cite{T-matrix-satisfy}
\begin{equation}\label{solve}
\pmb \pi  =\left( {{\pi_0},{\pi_1} \cdots {\pi_L}} \right)^{T}= \mathbf{Z}^{T} \pmb\pi.
\end{equation}
Note that this unique solution $\pmb \pi $ is the discrete distribution of the source residual energy and it can be solved from (\ref{solve}) as \cite{T-matrix-satisfy}
\begin{equation}\label{app-energy-distri}
\pmb\pi  = {\left( {{\mathbf{Z}^T} - \mathbf{I} + \mathbf{B} }\right)^{ - 1}}\mathbf{b},
\end{equation}
where ${\mathbf{B}_{i,j}}=1, \forall i, j$ and $\mathbf{b}={(1,1, \cdots ,1)^T}$. With the discrete distribution of the source residual energy, we now can evaluate the average throughput of the considered system in the following subsection.

\subsection{Throughput analysis of the DTS protocol}
For the considered system, the average throughput can be calculated as the product of the fixed rate $R$ and the probability that the system can perform outage-free IT operation. Note that in the DTS protocol, outage only happens when EH phase is performed and successive transmission happens in all the IT blocks. Jointly considering the principle of the DTS protocol given in (8), we can achieve that the probability of the outage-free IT operation is actually equivalent to that of the event ${\varepsilon _T} \le {\varepsilon _R}$. Utilizing the distribution of the source residual energy $\pmb \pi$ evaluated in (\ref{app-energy-distri}), the average throughput of the DTS protocol can be characterized as
\begin{equation}\label{R2}
\begin{split}
{\Phi _{DTS}}
&=R\Pr \left\{ {\varepsilon _T} \le {\varepsilon _R} \right\}\\
& = R\sum\limits_{i = 0}^L {\Pr \bigg \{  \big( {{\varepsilon _R} = iC/L}\big) \cap \big( {\varepsilon _T} \le iC/L \big) \bigg \} }    \\
&  = R \sum\limits_{i = 0}^L { {\pi _i} \Pr \left\{ {G \ge {{\upsilon {N_0}L} \over {iC}}} \right\}} \\
&  = R \sum\limits_{i = 1}^L {{\pi _i} \left[ {\exp \left( { - {\upsilon {N_0} L \over i C\Omega }} \right)\sum\limits_{n = 0}^{N - 1} {{{{{\left( {{\upsilon {N_0} L\over i C\Omega }} \right)}^n}} \over {n!}}}}\right]} .
\end{split}
\end{equation}

\begin{remark}
First of all, from (\ref{R2}), it is clear that the average throughput increases as the number of antennas $N$ increases. Furthermore, we can verify that the overall function $\exp \left( { - {x} } \right)\sum\limits_{n = 0}^{N - 1} {{{{{\left( {{x }} \right)}^n}} \over {n!}}}$ within the square brackets is an decreasing function of $x$ when $x\in \left[ {0,\infty } \right]$. Thus, either the decrease of the noise power $N_0$ or the increase of the channel variance $\Omega$ will increase the average throughput of the DTS protocol as expected. Increasing bit rate $R$ contributes oppositely in the average throughput of the system. On one hand, it increases the average throughput since it is a product term in (\ref{R2}). On the other hand, at the same time it increases the value of $\upsilon  = {2^R} - 1$, which reduces the function within the square brackets as analyzed above and  decreases the average throughput oppositely. Thus, we can conclude that when the other system parameters are fixed, there should exist an optimal bit rate $R$ such that the average throughput of the DTS protocol is maximized. Due to the complex structure of the distribution $\pmb \pi $ in (\ref{R2}), it is difficult to further derive a closed-form expression for the optimal bit rate. However, the optimal value $R$ can be readily obtained based on the derived expression (19) via numerical methods.
\end{remark}

\section{Numerical Results}
In this section, we present some numerical results to illustrate and validate the above theoretical analyses. A comparison of the DTS protocol with the existing HTT protocol is also provided. In order to capture the effect of path-loss on the system performance, we set $\Omega  = {10^{ - 3}}{d^{-\alpha} }$, where $d$ is the distance between the AP and source, $\alpha  \in \left[ {2,5} \right]$ is the path-loss exponent, the coefficient $10^{ - 3}$ corresponds to the propagation assumption of a 30dB average signal power attenuation per 1 meter (m) \cite{harvest-then-transmit}. In the following simulations, we set the distance $d=10$m, the path-loss factor $\alpha=2$, the noise power ${N_0} =  - 90$dBm, and the energy conversion efficiency $\eta  = 0.5$.

\begin{figure}
\centering \scalebox{0.38}{\includegraphics{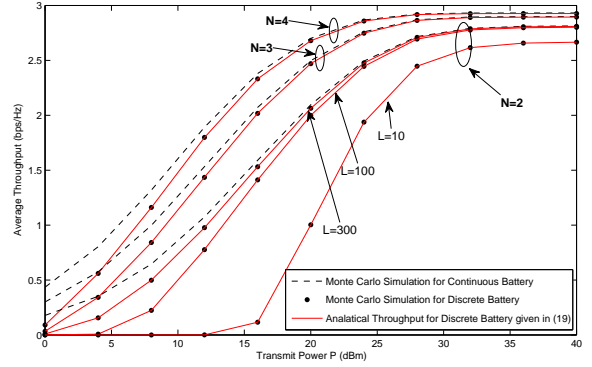}}
\caption{Average throughput versus transmit power for different antenna numbers $N=[2,3,4]$ and levels $L=[10, 100, 300]$ with fixed capacity $C= 2\times 10^{-5}$ and bit rate $R=3$. }\label{fig:1}
\end{figure}
We first compare the analytical throughput derived in above sections with the Monte Carlo simulation results. The average throughput of the DTS protocol versus AP transmit power for different antenna numbers and  discrete battery levels is shown in Fig. \ref{fig:1}. In Fig. \ref{fig:1}, we also present the average throughput of the DTS protocol for an ideal continuous battery (i.e., $L \to \infty  $) as a reference. We can see from this figure that the theoretical analysis given in (19) agree well with its corresponding Monte Carlo simulation, which validates our throughput analysis in Sec. III. Moreover, it can be seen that as the discrete level $L$ increases, the average throughput of the system increases and approaches the ideal case that the source has a continuous battery. This is understandable since for the discrete battery case, the source battery is charged to energy level $i$ only when the harvested energy is greater than level $i$. Part of the harvested energy is wasted through this discretization and less amount of energy is wasted when the interval between adjacent energy levels is reduced. Furthermore, from the results given in Fig. \ref{fig:1}, the average throughput derived in (\ref{R2}) can be used to approximate the throughput for an ideal continuous source battery when $L$ is large enough. Furthermore, when $L$ is fixed (e.g., $L = 300$ in Fig. 3), the larger the AP transmit power, the more accurate the approximation. This is because that for the same degree of precision, smaller intervals are required to model the source battery when the harvested energy is low.

\begin{figure}
\centering \scalebox{0.38}{\includegraphics{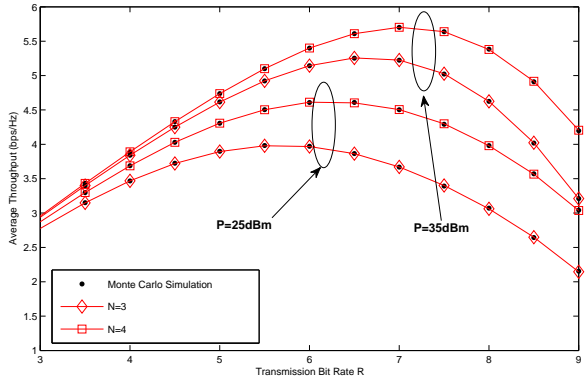}}
\caption{Average throughput versus transmission bit rate $R$ for $C=2 \times 10^{-5}$, $L=200$ and different antenna numbers $N=[3,4]$. }\label{fig:3}
\end{figure}

The average throughput of the DTS protocol versus different bit rate $R$ is depicted in Fig.\ref{fig:3}. It can be observed that there exists an optimal bit rate $R$ which maximizes the average throughput of the DTS protocol in all considered cases. This validates our discussion given in Remark 1. In addition, as the transmit power increases, the optimal bit rate $R$ shifts to the right. This is because a larger transmit power can support a higher bit rate under the same outage probability. Furthermore, the optimal bit rate $R$ reduces as the antenna numbers decreases.

\begin{figure}
\centering
\subfigure []
  {\scalebox{0.39}{\includegraphics {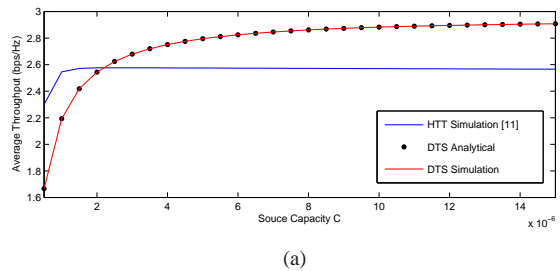}
  \label{a}}}
\hfil
\subfigure []
  {\scalebox{0.38}{\includegraphics {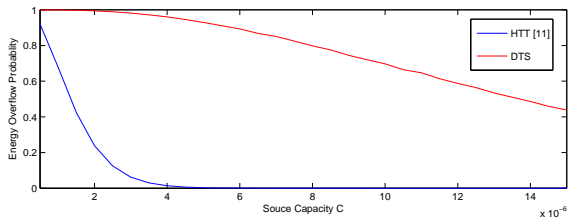}
\label{b}}}
\caption{Average throughput and energy overflow probability versus source capacity for DTS and HTT protocols where $R=3$, $N=3$, $L=300$ and $P=30$dBm.
\label{c}}
\end{figure}

Fig. 5 compares the average throughput of the DTS protocol and the HTT protocol for different source battery capacities. The energy harvesting time for the HTT protocol is set to its optimal value to maximize its average throughput \cite{optimal}. First of all, we can observe that the average throughput of the DTS protocol grows monotonically as the capacity increases. In contrast, the average throughput of the HTT protocol seems independent of the source capacity as it almost remains the same when the source capacity is large. This phenomenon can be understood as follows. The amount of harvested energy during each EH operation in the DTS protocol could be much larger than that in the HTT protocol. This is because the DTS protocol uses entire block to harvest energy,  while the HTT protocol only uses a small fraction. We thus can deduce that when the source capacity is limited, energy overflow happens in higher probability for the DTS protocol than the HTT protocol. Fig. 5 (b) have verified this deduction. This also explains why the value of battery capacity can influence the average throughput of the DTS protocol significantly, as shown in  Fig. 5 (a). Finally, we can also conclude from Fig. 5 that DTS protocol considerably outperforms the HTT protocol when the source capacity is large.

\section{Conclusions}
In this paper, we proposed a discrete time-switching protocol for multi-antenna wireless-powered communication networks. We modelled the energy accumulation process of the discrete-level source battery as a discrete Markov chain. An exact expression of the average throughput of the DTS protocol was derived over Rayleigh fading channels. All the theoretical analysis was then verified by the numerical simulation. Results showed that the DTS protocol considerably outperforms the HTT protocol when the source capacity is~large.
\appendices

\ifCLASSOPTIONcaptionsoff
  \newpage
\fi

\bibliographystyle{IEEEtran}
\bibliography{References}

\end{document}